\documentclass[aps,prl,twocolumn,groupedaddress,dvips]{revtex4}
\usepackage[dvips]{graphicx}
\begin{document}
%\preprint{}
\title{Spin-Based Magnetofingerprints and Dephasing in Strongly Disordered Au-Nanobridges}
\author{A. Anaya, M. Bowman, A. L. Korotkov, and D. Davidovi\'c}
%\email[]{dragomir.davidovic@physics.gatech.edu}
%\thanks{}
\affiliation{Georgia Institute of Technology, Atlanta, GA 30332}
%\collaboration{} FG^{as}
\date{\today}
\begin{abstract}
We investigate quantum interference effects with magnetic field
(magnetofingerprints) in strongly disordered Au-nanobridges. The
magnetofingerprints are unconventional because they are caused by
the Zeeman effect, not by the Aharonov-Bohm effect. These
spin-based magnetofingerprints are equivalent to the Ericson's
fluctuations (the fluctuations in electron transmission
probability with electron energy). We present
 a model based on the
Landauer-Buttiker formalism that describes the data. We show that
the dephasing time $\tau_\phi (E,T)$ of electrons at temperature
$T$ and energy $E$ above the Fermi level can be obtained from the
correlation magnetic field. In samples with localization length
comparable to sample size, $h/\tau_\phi (E,T) \approx E$, for
$E\gg k_B T$, which shows that the Fermi liquid description of
electron transport breaks down at length scale comparable to the
localization length.
\end{abstract}

%\keywords{}
\maketitle

Studies of electron transport through very\ disordered nanometer
scale materials are important for development of nano-devices. For
example, nanometer scale bridges between noble metals have
recently been fabricated using electrodeposition and
electromigration of noble
metals.~\cite{morpurgo,park,li,yu,yu1,bowman,boussaad,castle} In
certain cases, these nano-bridges can be highly disordered. The
disorder can cause Coulomb Blockade or strong suppression in the
density of states at low temperatures.~\cite{yu,yu1,bowman} In
this paper, the disorder is defined to be strong if the sample
conductance at zero bias voltage and low temperatures is much
smaller than the conductance at room temperature.

In our recent papers, we have reported on Coulomb-blockade and
universal conductance fluctuations in strongly-disordered Au
nanobridges formed by electromigration at room temperature in high
vacuum.~\cite{anaya,bowman,anaya1} The disorder depends on growth
conditions such as partial water vapor pressure in high vacuum and
the applied voltage. The Drude mean free path in the nano-bridge
material was estimated to be $\sim 0.01\lambda_F$, showing that
the disorder was significant. The origin of the disorder was
attributed to intermixing between the metal and $H_2O$
molecules.~\cite{bowman} We hypothesized that the disorder was
granular, because Au did not alloy with $H_2O$ (so the disorder
could not be amorphous).

Recently, granular disorder has been directly observed in
Au-nanoelectrodes grown by electrochemistry.~\cite{nieto,grose} In
particular, an electrochemical growth process at certain gate
voltages created a granular material consisting of Au clusters
with oxidized surface. Granular disorder can cause characteristics
that resemble  single-electron transistors, and researchers were
cautioned about interpreting transistor effects in
molecular-electronics experiments.~\cite{grose}

The goal of this paper is to explain quantum interference effects
in strongly disordered Au nano-bridges at low temperatures, using
a model of granular disorder. This paper is a follow-up to our
recent letter~\cite{anaya1} in which we demonstrated phase
coherent phenomena in strongly disordered Au nano-bridges. We
reported quantum interference effects at the Fermi level similar
to universal conductance fluctuations in weakly disordered
metals.~\cite{washburn}

In the first part of the paper, we describe experimental results.
The main observation is a linear structure in conductance versus
bias voltage and magnetic field at low temperature. The structure
demonstrates that electron interference effects with magnetic
field are spin-based, in contrast to the similar effects in weakly
disordered metals,
 which are caused by the Aharonov-Bohm effect on the phase of the electron wavefunction.~\cite{washburn}

Spin-based interference effects were explained by the granular
disorder model.~\cite{anaya1} In this paper we present an in-depth
discussion of the granular nano-bridge model. We introduce
diffusion and resistor network approaches to granular conduction,
and demonstrate the equivalence of the approaches, analogous to
the Einstein relation in homogeneous systems.

We show that the criterion to observe spin based
magnetofingerprints is that the Drude mean free path ($l_D$) be
smaller than the Fermi wavelength ($\lambda_F$). In homogeneous
systems, $l_D <\lambda_F$ would imply that the system is an
insulator at zero temperature.~\cite{imry} However, in a granular
system, $l_D$ can be much smaller than $ \lambda_F$ even in the
metallic state, because $l_D$ is not the same as the electron
scattering length.~\cite{abeles,imry} This is why we can observe
electron transport  over relatively long distances ($>25nm$) at
the Fermi level. We present numerical simulations of quantum
interference effects in a regime where $l_D <\lambda_F$, using a
diffusive description of hopping conduction.

Finally, we present bias-voltage dependence of conductance
fluctuations, which has not been discussed yet. We obtain bias
voltage dependence of the correlation magnetic field, and show
that this dependence displays the energy dependence of the
dephasing time. In samples with resistance $\sim R_Q=h/e^2$, we
find that the energy uncertainty of electrons at energy $E$ above
the Fermi level is the same as $E$. This shows that the
Fermi-Liquid description of electron transport breaks down when
the sample size is comparable to the localization length.

\section{Experiment Description}

In this section we describe sample fabrication and demonstrate
spin-based quantum interference effects with magnetic field at low
temperatures.

\subsection{Sample fabrication}

\begin{figure}%[p]
\includegraphics[width=0.47\textwidth]{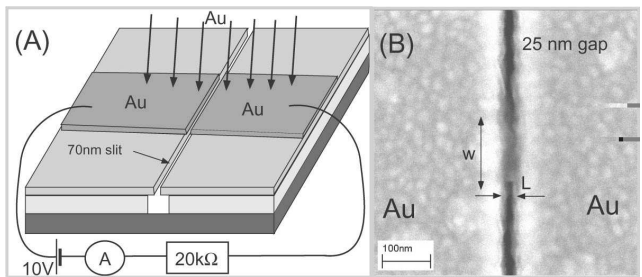}
\caption{A. Sketch of sample fabrication. B. Au nano-bridge with
resistance of $10k\Omega$.} \label{fabrication}
\end{figure}

Nano-bridge fabrication process was discussed in detail in our
prior publications.~\cite{anaya,bowman,anaya1} In summary, samples
were created by deposition of Au atoms in high vacuum ($\sim
10^{-7}Torr$) over two bulk Au films separated by a $\sim 70$nm
slit, as sketched in Fig.~\ref{fabrication}-A. A finite bias
voltage is applied between the films and the resulting current is
measured in situ. The deposition of Au is stopped as soon as a
finite current is detected. In this way, only one contact is
created.

The shape and the resistance of the contacts depends on growth
conditions, such as bias voltage and background pressure. For
example, we demonstrated that the resistance of the contact at
high bias voltage can be increased reversibly by several orders of
magnitude by increasing the partial water vapor pressure around
the contact.~\cite{korotkov} In general, the growth process is
quite complicated, because processes such as electromigration,
surface atom diffusion, and intermixing with water molecules
affect the contact shape and the disorder.~\cite{bowman}

The contacts selected in this work are grown at 10V using a
limiting resistor of $R=20k\Omega$.  The purpose of the limiting
resistor is to reduce the current as soon as the contact is
created, which prevents catastrophic electromigration processes.

One typical sample is shown in Fig.~\ref{fabrication}-B. The
resistance of the contact is approximately inversely proportional
to contact apparent width $w$.~\cite{bowman} Thus, the resistance
of the contact material is approximately uniformly distributed.
The disorder length scale (grain size) is much smaller than the
contact size $L\sim 25nm$.

In wide contacts ($w>50nm$), $R<R_Q=h/e^2$ and samples do not
display Coulomb Blockade at our lowest temperature ($15mK$). In
narrow contacts ($w<50nm$), $R> R_Q$ and samples exhibit Coulomb
Blockade at low temperatures.~\cite{bowman} Thus, in samples where
$w\sim 50nm$, the electron localization length is comparable to
$L$.

The resistivity of the material inside the nano-bridge was
estimated to be $\rho\sim 10^5\mu\Omega cm$,~\cite{bowman} which
is larger than the maximum metallic resistivity of $10^3\mu\Omega
cm$.~\cite{mott} The corresponding Drude mean free path ($l_D$),
obtained from $\rho=mv_F/ne^2l_D$, is $l_D\sim 0.01\lambda_F$.

\subsection{Spin-Based Magnetofingerprints}

We present quantum interference effects with magnetic field in
devices with resistance $<R_Q$, because these devices do not
exhibit Coulomb Blockade at low temperatures, which makes it
possible to study conductance fluctuations at the Fermi level,
analogous to conductance fluctuations in weakly disordered
metals.~\cite{webb} We select two new devices with resistances
slightly below $R_Q$.

\begin{figure}%[p]
\includegraphics[width=0.47\textwidth]{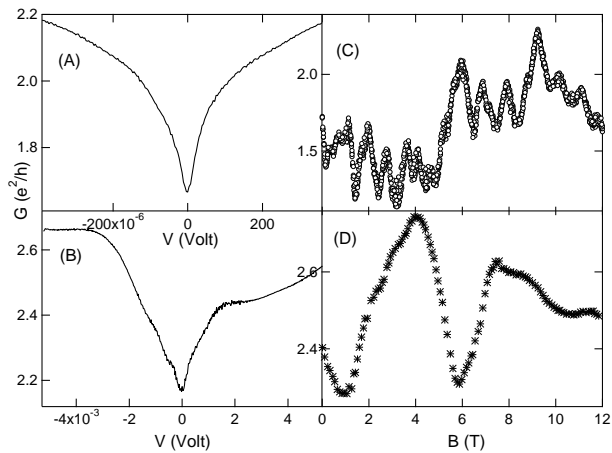}
\caption{A and B: Differential conductance versus bias voltage at
$T=0.015K$ in samples 1 and 2, respectively. C and D: Differential
conductance versus magnetic field at $T=0.015K$ in samples 1 and
2, respectively.} \label{zba}
\end{figure}

The I-V curves display a conductance suppression at zero bias
voltage and at low temperatures (zero-bias anomaly or ZBA). This
suppression is a Coulomb Blockade precursor.~\cite{anaya1}
Figs.~\ref{zba}-A and B display differential conductance $G=dI/dV$
versus $V$ in samples 1 and 2, respectively. $G$ is measured by
lock-in technique with an excitation voltage $3\mu V$. The ZBA was
explained in detail by the theory of broadened Coulomb Blockade in
a single electron transistor.~\cite{golubev}

Figs.~\ref{zba}-C and D display differential conductance versus
magnetic field of samples 1 and 2 at zero DC-bias at base
temperature. The conductance exhibits reproducible fluctuations
with field. The fluctuation amplitude close to the universal value
($\sim e^2/h$). The fluctuations are sensitive to thermal cycling
to room temperature, which demonstrates sensitivity of the
fluctuations to impurity configurations. Thus, the fluctuations
are analogous to the well known magnetofingerprints in mesoscopic
physics.

\begin{figure}%[p]
\includegraphics[width=0.47\textwidth]{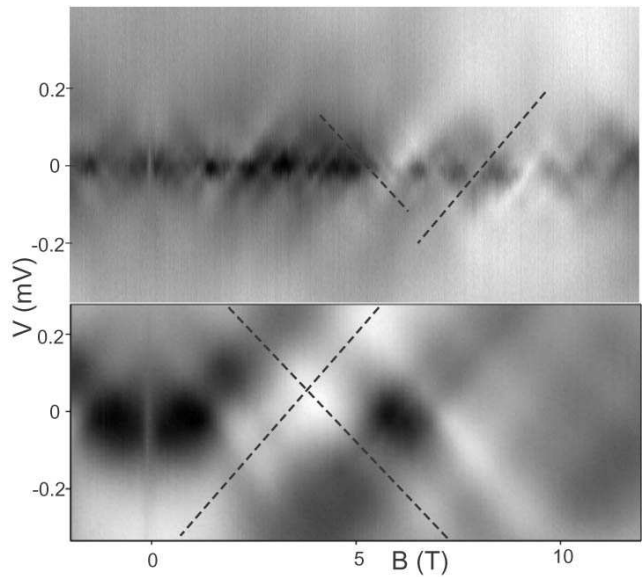}
\caption{A and B: Differential conductance versus magnetic field
and bias voltage at $T=0.015K$ in samples 1 and 2, respectively.}
\label{spin}
\end{figure}

The magnetofingerprints in our samples are caused by the Zeeman
effect. This is demonstrated in Fig.~\ref{spin}, which represents
$G$ versus magnetic field and bias voltage. The figure displays a
linear structure, which is highlighted by dashed lines of
mathematical form $eV\pm 2\mu_B B=const$. We investigated a set of
10 samples with resistances in range $(5k\Omega,15k\Omega)$ at
milli-Kelvin temperature. The linear structure was found in every
one of these samples. Since the slopes of the lines correspond to
the Zeeman splitting, this confirms that the fluctuations with
magnetic field are spin-based. By contrast, conductance
fluctuations in weakly disordered metals are based on the
Aharonov-Bohm effect and $G$ versus $V$ and $G$ versus $B$ are
parametrically inequivalent.

 In the
next 2 sections of this paper, we will demonstrate that
magnetofingerprints become spin based when the Drude mean free
path becomes shorter than the Fermi wavelength. If the conducting
material were homogeneous, this condition would not be physical
because
 the Drude mean free path and the electron scattering
length would be the same. But, if the system is not homogeneous,
the Drude mean free path and the electron scattering length will
not be the same and $l_D$ can be $\ll\lambda_F$, even if the
conducting material is a metal at $T=0$.

\section{Granular Nano-bridge Model}

In this section we introduce a model of electron transport through
the nano-bridge. The model assumes that the disorder is granular
with the grain size much smaller than $L$. The grain diameter
($D$) is certainly smaller than our imaging resolution of roughly
$5 nm$.
%The resolution is not sufficient to determine $D$.

As discussed in the introduction, amorphous disorder is ruled out
because Au does not alloy with the impurities that are responsible
for disorder ($H_2O$). Granular disorder has been recently
confirmed in $Au$ samples created by
electro-chemistry.~\cite{nieto,grose} It has been shown that large
electric fields near Au surfaces induce formation of Au
nanoparticles of diameter of few nanometers, which assemble into a
granular structure.~\cite{grose}

Theoretically, granular metals are similar to Fermi-Liquids. They
are also known as Granular Fermi Liquids.~\cite{beloborodov2} It
has long been recognized that the Drude mean free path of a
granular metal $l_D$ can be much shorter than the electron Fermi
wavelength,~\cite{abeles,imry} because $l_D$ and the electron
scattering length $l$ are not the same.

\begin{figure}%[p]
\includegraphics[width=0.47\textwidth]{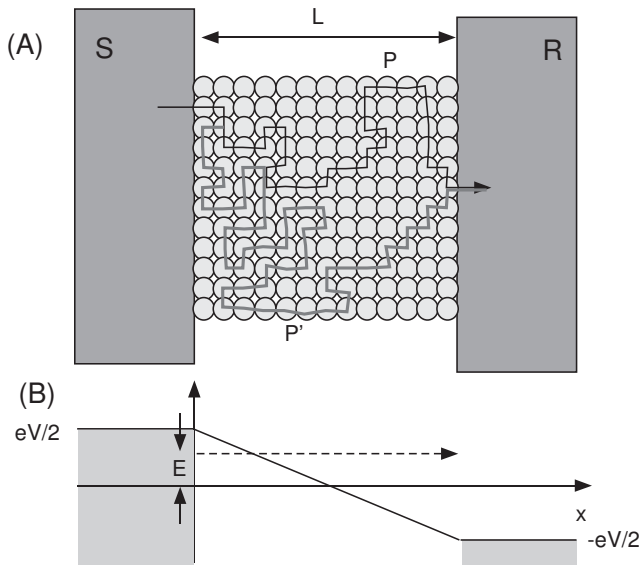}
\caption{A. Mesoscopic granular conductor connected between
reservoirs S and R. B. Quasi-Fermi level versus position when $S$
is a source.}~\label{model}
\end{figure}

Consider a granular bridge of length $L\gg D$  in
Fig.~\ref{model}-A. The resistance of the bridge can be obtained
using a 3D-resistor network model, because, by definition, the
resistance between the grains ($R_G$) is larger than the
resistance inside the grains (otherwise, the system would be
homogeneous). The resistance of the network is $R=R_GDL/A$, where
$A$ is the sample cross-section area. The resistivity is obtained
from $R=\rho L /A$, so $\rho=R_G D$.

In homogeneous systems, one can apply the Drude formula for
resistivity $\rho=\frac{mv_F}{ne^2l}$, where $l$ is the electron
scattering length. In granular systems, however, the Drude formula
is not valid. But, one can still apply the Drude formula
$\rho=\frac{mv_F}{ne^2l_D}$, where $l_D$ will be referred to as
the nominal Drude mean free path. In a resistor network,

\begin{equation}
l_D=\frac{mv_F}{ne^2 R_G D}. \label{drude}
\end{equation}
$l_D$ is not the same as electron scattering length $l$ ($l\sim D$
for ballistic grains). One can change $l_D$ without changing $l$,
by changing $R_G$.

We expect that the metal insulator transition occurs  roughly when
$R_G\sim zR_Q$, where $z$ is the coordination number ($z=6$ in our
model). This condition is reasonable, because if the total
resistance between a grain and its neighbors exceeds $R_Q$, then
electron transport will be blocked by Coulomb-blockade and the
system will be an insulator. Theoretical value for  $R_G$ at the
metal-insulator transition is similar to our
estimate.~\cite{beloborodov1}

%In our nano-bridges $\rho\sim 10^5\mu\Omega cm$. Substituting
%$\rho = R_G D$, we find that the nano-bridge material would be a
%metal if $D>6nm$ and it would be an insulator if $D<6nm$.

A granular metal will satisfy the condition $l_D<\lambda_F$ if the
inter-grain resistance is in interval
$R_Q\frac{\lambda_F}{D}<R_G<zR_Q$. At the metal-insulator
transition, $l_D$ has a critical value $\lambda_F^2/(zD)$, which
can be several orders of magnitude shorter than $\lambda_F$.

In this paper, we investigate mesoscopic electron transport in
this regime ($l_D <\lambda_F$). We will show that $l_D$ plays a
crucial role in determining the origin of mesoscopic fluctuations
with magnetic field.

Next, we discuss electron diffusion through granular systems. We
assume that the average electron dwell time on any given grain is
$\tau_D$. After time $t\gg\tau_D$, there are roughly
$\frac{t}{\tau_D}$ hops between neighboring grains. If we assume
that the hops are uncorrelated, then electron motion is diffusive,
and the diffusion length is $L=
D\sqrt{\frac{t}{3\tau_D}}=\sqrt{\frac{D^2}{3\tau_D}t}$. Thus, the
diffusion coefficient is $D^{dif}=\frac{D^2}{3\tau_D}$.

In a homogeneous metal, the diffusion coefficient is $D^{dif}=v_F
l/3$. In a granular metal, this relationship is not valid. One can
still apply this formula, $D^{dif}=v_F l_{dif}/3$, but $l_{dif}$
will not be the same as $l$ and it will be referred to as the
nominal diffusion mean free path. It follows that

 \[l_{dif}=\frac{D^2}{v_F\tau_D}.\]

Next, we investigate the relationship between the nominal Drude
mean free path and the nominal diffusion mean free path. Let us
assume that $l_{dif}=l_D$.
 Then
$\tau_D=\frac{D^3R_Gne^2}{mv_F^2}$, and after some algebra it can
be shown that this is equivalent to

\begin{equation}
\tau_D=\frac{h}{\delta}\frac{R_G}{R_Q}. \label{dwell}
\end{equation}
where $\delta$ is the level spacing inside the grains and
$R_Q=h/e^2$ is the resistance quantum.

This is the well known formula for the dwell time of an electron
on a quantum dot connected to leads via resistance $R_G$.
Eq.~\ref{dwell} is undoubtedly correct, so it follows that
$l_{dif}=l_D$. Thus, when examining electron transport over
distances much larger than grain diameter, one can use a diffusion
model with $l_D$ obtained from the resistor-network
model(Eq.~\ref{drude}). Formula~\ref{dwell} is essentially the
Einstein relation between diffusion and conductivity of a granular
metal.

\section{Magnetofingerprints when $l_D<\lambda_F$}

In this section, we explain the spin-based origin of conductance
fluctuations ($G(B,V,T)$) and discuss linear correlations in
differential conductance with magnetic field $B$ and bias voltage
$V$. First we examine fluctuations in differential conductance at
$T=0$. Finite temperature effects are discussed in the appendix.

We will neglect the effects of spin-orbit scattering. In our prior
publication, we demonstrated that spin-orbit scattering is
suppressed because of the quantization of energy levels inside the
grains.~\cite{anaya1}

\subsection{Magnetofingerprints as the Ericson's fluctuations.}

First we investigate conductance fluctuations with field at zero
DC bias voltage. We find the conductance contribution from
electrons with spin-up, $G_\uparrow (B,0,0)$. The differential
conductance is given by the Landauer-Buttiker
formula,~\cite{datta}
\begin{equation}
G_\uparrow(B,0,0)=\frac{e^2}{h}T(E,B),\label{landauer}
\end{equation}
where $T(E,B)$ is the transmission probability of electrons with
energy $E$ in magnetic field $B$. By our convention, the energy in
this equation is measured relative to the Fermi energy $E_F$, i.
e. $E=0$ for electrons with energy $E_F$.

The magnetic field dependence in $T(E,B)$ originates from the
Aharonov-Bohm effect on the phase of the electron wavefunctions.
We will neglect the Aharonov-Bohm effect, so that the dependence
$T(E,B)$ on the second argument can be omitted, i.e.
$T(E,B)=T(E)$. This approximation is valid if the applied magnetic
flux through the sample is much smaller than the flux quantum.
Later in the text, we discuss the validity of this  assumption.

At zero bias voltage, $E$ is just the kinetic energy. The
dependence of the transmission probability on $E$ arises because
electron wavelength decreases with kinetic energy. This causes
mesoscopic fluctuations known as the Ericson's
fluctuations.~\cite{ericson} The Ericson's fluctuations in
mesoscopic conductors have not yet been measured directly.

Without Zeeman splitting, $E$ in Eq.~\ref{landauer} would be equal
to $0$. However, with Zeeman splitting included, the kinetic
energy of spin-up electrons at the Fermi level increases with
magnetic field. In particular, at the Fermi level the sum of the
kinetic energy and the Zeeman energy equals $E_F$. So $E=g\mu_B
B/2$, where $g=2$ is the g-factor, and $\mu_B$ is the Bohr
magnetron. Consequently, Eq.~\ref{landauer} becomes
\begin{equation}
G_\uparrow(B,0,0)=\frac{e^2}{h}T (\mu_B B).
\end{equation}
Taking into account electrons with spin down, the linear
(zero-bias) conductance becomes
\begin{equation}
G(B,0,0)=\frac{e^2}{h}\left(T(\mu_B B)+T(-\mu_B B)\right).
\label{landauer1}
\end{equation}

This equation demonstrates that magnetofingerprints are equivalent
to the Ericson's fluctuations. This behavior is possible only
because we neglected the Aharonov-Bohm effect. The correlation
energy $E_C$ is related to the correlation field $B_C$: $E_C=\mu_B
B_C$. At $T=0$ and $V=0$, $E_C$ is given by the Thouless energy
$E_{Th}=\frac{hD^{dif}}{L^2}$.~\cite{washburn}

\subsection{Conductance Fluctuations with Bias Voltage and the Ericson's fluctuations}

Now we consider the case when bias voltage $V$ is applied between
the reservoirs. We choose a symmetric bias, in which the
electrostatic potential of reservoir S is $V/2$ and the
electrostatic potential of reservoir L is $-V/2$. We assume that
reservoir $S$ is a source ($V<0$). One would naively expect that
the fluctuations of $G(B,V,0)$ with $V$ are similar to the
Ericson's fluctuations, because increasing $V$ increases the
kinetic energy of electrons that participate in transport.

We will assume that the energy is conserved in electron transport,
i. e. the electron transport is horizontal, as sketched in
Fig.~\ref{model}-B. At $V\neq 0$, the quasi-Fermi level inside the
sample varies with position. As a result, the kinetic energy
varies with $x$, so strictly speaking $V$ fluctuations will not be
equivalent to the Ericson's fluctuations.

The transmission probability depends on both electron energy and
bias voltage. In further discussion, $T(E,V)$ is the probability
of transmission from S to R for electrons with energy $E$ at bias
$V$. $E$ is measured relative to the average Fermi level of the
two reservoirs, as shown in Fig.~\ref{model}-B.

 In
absence of inelastic electron scattering, the current through the
sample is given by the Landauer-Buttiker formula,~\cite{datta}:
\begin{widetext}
\begin{equation}
I(B,V,0)=\frac{|e|}{h} \sum_{\alpha=\pm 1}\int T(E+\alpha\mu_B
B,V)\left(f(E -\frac{eV}{2})-f(E+\frac{eV}{2})\right)dE.
\label{landauerv}
\end{equation}
\end{widetext}
where $f(E)=1/(exp(E/k_BT)+1)$ is the Fermi function at $T=0$ and
summation over $\alpha$ indicates sum over electron spin
polarization.

In the Landauer formalism, electron transport is horizontal, so
the kinetic energy increases with position ($x$), because the
quasi-Fermi level decreases with x. The quasi-fermi level is
defined as usual, as the position of the Fermi level that the
electrons would assume if they were in local equilibrium (this
would be the case if electron-phonon relaxation rate were much
larger than the electron transport rate).

In good metals, the quasi-fermi level is very close to the
electrostatic potential energy $e\phi(x)$, where $\phi(x)$ is the
electrostatic potential and $e$ is the electron charge ($e<0$). We
assume that the electric field inside the sample is uniform, so
$e\phi (x)=eV/2-eVx/L$ for $0<x<L$.

The transmission coefficient can be expressed as
\begin{equation}
T(E,V)=\sum T_{m,n}(E,V),
\end{equation}
where the sum is taken over incoming channels ($m$) in the left
reservoir and all the outgoing channels ($n$) in the right
reservoir, and $T_{m,n}(E,V)$ is the transmission probability from
channel $m$ into channel $n$. $T_{m,n}(E,V)$ can be expressed in
terms of the probability amplitude $t_{m,n}(E,V)$ for transfer
from channel m to channel n, $T_{m,n}(E,V)=|t_{m,n}(E,V)|^2$.
Finally, the probability amplitude is equal to the sum over
probability amplitudes of various Feynman paths connecting the
reservoirs,~\cite{datta}
\begin{equation}
t_{m,n}(E,V)\sim\sum_{P}e^{i\phi_P(E,V)},
\end{equation}
where $P$ indicates an electron  path and $\phi_P(E,V)$ is the
phase accumulated by an electron injected from reservoir S at
energy $E$ along this path.

In the semiclassical approximation,
$\phi_P(E,V)=\phi_P+\hbar^{-1}\int_0^{t_P} \delta E(t) dt$, where
$\delta E(t)$ is the kinetic energy of an electron on path $P$ at
moment $t$, and $\phi_P=\phi_P(0,0)$. So, $\delta
E(t)=E-e\phi\left[x(t)\right]$ and
\[
t_{m,n}(E,V)\sim\sum_{P}e^{i\phi_P-iEt_P/\hbar+ieV\int_0^{t_P}\left(1/2-x(t)/L\right)dt/\hbar},
\]

This expression can be rewritten as
\begin{equation}
t_{m,n}(E,V)\sim\sum_{P}e^{i\phi_P-iEt_P/\hbar+ieVt_P\Sigma(P)/\hbar},
\label{phases}
\end{equation}
where $\Sigma
(P)=\frac{1}{t_P}\int_0^{t_P}\left[1/2-x(t)/L\right]dt$.

The dependencies of $T(E,V)$ on $E$ and $V$ are not equivalent,
because phase contributions $Et_P/\hbar$ and $\Sigma (P)
eVt_P/\hbar$ are not directly proportional, since $\Sigma(P)$
fluctuates among different paths. If $P$ is chosen so that an
electron spends roughly the same time at different locations $x$,
as sketched in Fig.~\ref{model}-A, then $\Sigma (P)\approx 0$. If
an electron spends significantly more time closer to one reservoir
than to the other reservoir, which is sketched by path $P'$ in
Fig.~\ref{model}-A, then $\Sigma(P')\approx 1$.

We determined the statistical distribution of $\Sigma (P)$ among
different paths. To this end, we generated random-walks on 3D
grids connecting two reservoirs. We find that the distribution of
$\Sigma$ is a Gaussian with a probability density
$f(\Sigma)=(2\pi\sigma^2)^{-1/2}\exp(-\Sigma^2/(2\sigma^2))$,
where $\sigma=0.167$. The value of $\sigma$ is important, because
it will set the voltage range where V-fluctuations and the
Ericson's fluctuations are equivalent.

Next we obtain the differential conductance $G(B,V,T=0)=\partial
I(B,V,T)/\partial V$,
\begin{widetext}
\begin{eqnarray}
G(B,V,0)=\sum_{\alpha=\pm 1}\frac{e^2}{2h}\int dE\cdot
T(E+\alpha\mu_BB,V)\left(-f'(E-\frac{eV}{2})-f'(E+\frac{eV}{2})\right)
\nonumber \\ +\frac{|e|}{h}\sum_{\alpha=\pm 1}\int dE
\frac{\partial T(E+\alpha\mu_B B,V)}{\partial V}
(f(E-\frac{eV}{2})-f(E+\frac{eV}{2}) ).\label{conductance}
\end{eqnarray}

Substituting the Fermi function at $T=0$, we find

%\begin{equation}
%G_{\uparrow}(B,V,0)=\frac{e^2}{2h} \left[
%T(\frac{|e|V}{2}+\mu_BB,V)+T(-\frac{|e|V}{2}+\mu_BB,V)\right] +
%\frac{e}{h}\int_{-|e|V/2}^{|e|V/2} dE \frac{\partial
%T(E+\mu_BB,V)}{\partial V}. \label{conductance0}
%\end{equation}
%The total conductance is $G=G_\uparrow+G_\downarrow$,
\begin{equation}
G(B,V,T=0)=\frac{e^2}{2h}\sum_{\alpha,\alpha'=\pm 1} T(\alpha\mu_B
B+ \alpha'\frac{eV}{2},V)+ \frac{|e|}{h}\sum_{\alpha =\pm
1}\int_{-eV/2}^{eV/2} dE \frac{\partial T(E+\alpha\mu_B B
,V)}{\partial V} \label{conductance0}
\end{equation}
\end{widetext}

The Ericson's fluctuations are represented by the dependencies
$T(\alpha eV/2+\alpha'\mu_B B,V)$ on the first argument. If both
$B$ and $V$ are varied simultaneously so that $eV/2\pm\mu_B
B=const$, then two out of the four terms in the sum on the right
hand side of Eq.~\ref{conductance0} will not fluctuate, which will
lead to a linear structure in $G(B,V,0)$ similar to our
experimental observations.

The characteristic scales for the dependence of $T(\alpha
eV/2+\alpha'\mu_B B,V)$ on the first argument are
\begin{equation}
 |e|V_C/2=\mu_B B_C=E_C=E_{Th}.
\label{corvoltage}
\end{equation}

The linear structure in $G(B,V,0)$ will be significant
 if we
can neglect the dependence $T(\alpha eV/2+\alpha'\mu_B B,V)$ on
the second argument ($V$) and if we can neglect the dependence of
the integral on the right hand side of Eq.~\ref{conductance0} on
$B$ and $V$. Now we show that these two dependencies can indeed be
neglected, which explains the linear structure in our experiments.
In particular, we show that the correlation voltage $V_C^q$ for
these two dependencies is much larger than $V_C$.

The correlation energy ($E_C$) is roughly given by condition
$E_Ct_P/\hbar=\pi$ (Eq.~\ref{phases}) for a typical path P.
Similarly, the characteristic voltage ($V_C^q$) for the second
argument in $T(E,V)$ is obtained from condition $eV_C^q\sigma
t_P/\hbar=\pi$ (Eq.~\ref{phases}). It follows that  $V_C^q\approx
6E_C/|e|=3V_C$. The dependence $T(\alpha\mu_B B+\alpha' eV/2,V)$
on the second argument has three times larger correlation voltage
than that in the first argument, explaining the pronounced linear
structure in our experiments.

In conclusion, the conductance fluctuations with bias voltage are
equivalent to the Ericson's fluctuations and $E_C=|e|V_C/2$. The
equivalence persists only in voltage range $<3V_C$.

The integral in Eq.~\ref{conductance0} also leads to fluctuations
in conductance with $B$ and $V$. It has been predicted by Larkin
and and Khmelnitskii~\cite{larkin} that this integral causes an
increase in amplitude of conductance fluctuations at bias voltages
larger than $E_C/e$, by a factor of $\sim\sqrt{eV_C/E_C}$. The
increase has been observed in Ag wires.~\cite{terrier}

Now we discuss the characteristic $V$ and $B$ of the integral. The
integration over $E$ broadens the dependence of $\partial
T(E+\alpha\mu_B B,V)/\partial V$ on magnetic field, thereby
increasing the correlation field from $B_C$ to $eV/\mu_B$. Thus,
the correlation voltage of the integral is set by the V-dependence
in $T(E,V)$, and this correlation voltage is $V_C^q$, six times
larger than $V_C$. So, the integral will not obscure the linear
structure.

The dependencies of $T(\alpha\mu_B B+ \alpha'\frac{eV}{2},V)$ on
the second argument and the integral are responsible for the
asymmetry of conductance fluctuations around zero bias voltage.

\begin{figure}%[p]
\includegraphics[width=0.47\textwidth]{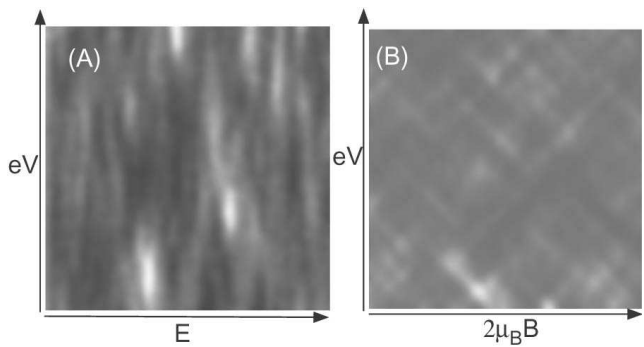}
\caption{A. Transmission probability $T(E,V)$ B. Differential
conductance $G(B,V)$.}\label{linessim}
\end{figure}

To investigate different terms in Eq.~\ref{conductance0}, we make
numerical simulations. We model our sample with a grid containing
$100\times 100\times 100$ grains, and generate 2000 random walks
across the sample. We calculate the phases $\phi(E,V)$ along these
random walks and obtain $T(E,V)$ from the transmission amplitudes
in Eq.~\ref{phases}. We assume that there is only one conducting
channel, for simplicity, and sum the transmission amplitude over
2000 randomly selected paths $P$. $T(E,V)$ is plotted in
Fig.~\ref{linessim}-A. The pattern is streched along $V$
direction, which shows that that correlation scale for $E$ ($E_C$)
is much smaller than correlation scale for $eV$ ($|e|V_C^Q$), in
agreement with the analysis above.

The differential conductance versus zeeman energy and bias voltage
are plotted in Fig.~\ref{linessim}-B. The linear structure is
clearly significant and similar to our experimental observations.
Note that there is a finite voltage range over which the lines in
$B-V$ parameter space are resolved. This voltage range is roughly
$3V_C$.

\subsection{Absence of the Aharonov-Bohm Effect}

We show that if $l_D<\lambda_F$, then quantum interference between
paths $P$ and $P'$ in Fig.~\ref{model}-A will be more sensitive to
the Zeeman effect than to the Aharonov-Bohm effect, which explains
the spin-origin of magnetofingerprints in our samples. The
characteristic magnetic field for the Aharonov-Bohm effect is
given by the field for a flux quantum $\Phi_0 =h/2e$ over sample
area, $B_{AB}=\Phi_0/L^2$.~\cite{washburn} Using
Eq.~\ref{corvoltage}, we find $B_C/B_{AB}\sim l_D/\lambda_F$, so
$B_C<B_{AB}$ if $l_D<\lambda_F$.

\section{Dephasing Effects}

\begin{figure}%[p]
\includegraphics[width=0.45\textwidth]{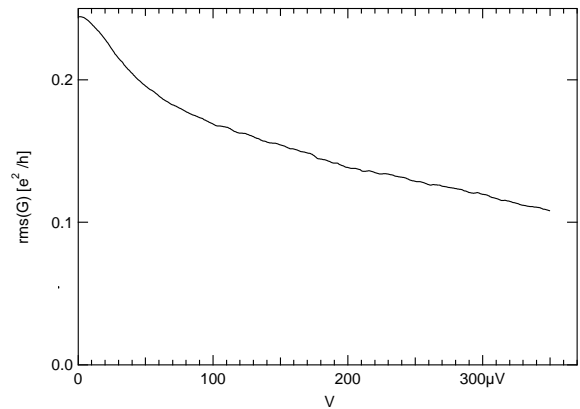}
\caption{Rms conductance versus bias voltage in sample 1 at
$T=0.015K$ .~\label{cfa}}
\end{figure}

In this section we discuss electron dephasing in our nano-bridges.
Samples can be divided into two groups. In the first group, the
conductance fluctuation amplitude and correlation scales $B_C$ and
$V_C$ have strong $T$ and $V$ dependence. For example, in sample 1
conductance fluctuation amplitude decreases rapidly with $|V|$.
Fig.~\ref{cfa} displays root-mean-square differential conductance
versus bias voltage at $T=0.015K$ in sample 1. The linear
structure is pronounced only near $V=0$, which indicates the the
correlation field increases with $|V|$. Eight out of ten samples
belong to the first sample group.

In the second group of samples, $B_C$ and $V_C$ at $V=0$ are
independent of $T$ below an experimentally accessible temperature.
In addition, base temperature values of $B_C$ and $V_C$ are
independent of $|V|$ below a certain $|V|$. This indicates that
thermal broadening and dephasing are weak at low $T$ and $|V|$, so
correlation scales are set by the Thouless energy,
$E_C=|eV_C/2|=\mu_B B_C=E_{Th}=hD^{diff}/L^2$. Sample 2 belongs to
this group. Two out of ten samples belong to the second sample
group. As discussed before, all ten samples have similar
resistance. The origin of different sample groups will be
discussed later in the section.

In the remainder of this section, we focus on the first group of
samples. The dephasing time ($\tau_\phi$) can be obtained from the
voltage dependence of the correlation field at base temperature.
This method for measuring $\tau_\phi$ is different from that in
standard mesoscopic physics of weakly disordered metals, where
$\tau_\phi$ can be measured from the temperature dependence of
$B_C$.~\cite{mohanty}

In our nano-bridges, magnetofingerprints are spin-based and $\mu
B_C$ is equal to the correlation energy.  In Appendix 1 we find
that if thermal broadening and dephasing are significant, then
\begin{equation} \mu_B B_C(V,T)\approx 2.9k_BT+h/\tau_\phi(E,T),
\label{bc}\end{equation} where $E=|eV/2|$. (The factor of 2.9 is
valid in a narrow temperature range $E_{Th}< 2.9k_BT <3E_{Th}$).
$\tau_\phi(E,T)$ is the dephasing time of electrons at energy $E$
above the Fermi level at temperature $T$.

It is not possible to obtain the temperature dependence of the
dephasing time $\tau_\phi$ from the temperature dependence
$B_C(0,T)$, because $B_C(0,T)$ is affected by both thermal
broadening and dephasing.  But, if $T$ is fixed, the energy
dependence of the dephasing time can be obtained from
Eq.~\ref{bc}.

\begin{figure}%[p]
\includegraphics[width=0.45\textwidth]{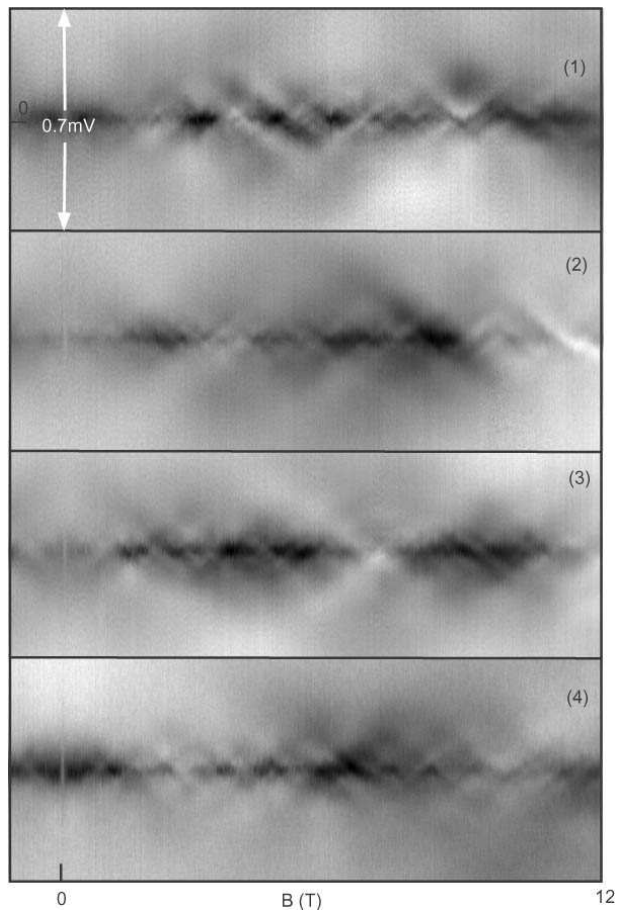}
\caption{(1)-(4): Differential conductance versus magnetic field
and bias voltage in sample 1 at $T=0.015K$, obtained in four
different thermal cycles.}\label{lines2}
\end{figure}

To increase the statistics, we thermally cycled sample 1 between
room temperature and base temperature 8 times. Fig.~\ref{lines2}
shows the linear structure in 4 successive cool-downs, indicating
that the data obtained in different thermal cycles are
statistically uncorrelated. Even though thermal cycling scrambles
the linear structure, the average conductance does not change with
thermal cycling.

\begin{figure}%[p]
\includegraphics[width=0.45\textwidth]{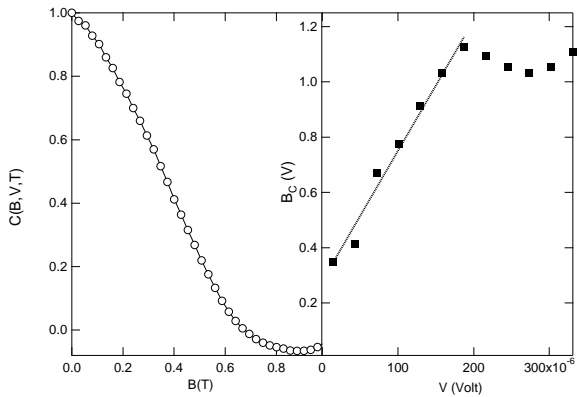}
\caption{A. Magnetic field autocorrelation function in sample 1 at
$T=0.015K$. B. Correlation magnetic field versus bias voltage at
$T=0.015K$.}\label{correl}
\end{figure}

$B_C(V,T)$ is calculated as the field at which the field
autocorrelation function ($C(B,V,T)$) decreases by 50$\%$, i.e.
$C(B_C(V,T),V,T)=0.5$. The field autocorrelation function at
temperature $T$ is found as
 \[
C(B,V,T)=\frac{\overline{G(B+B_0,V,T)G(B_0,V,T)}-\overline{G(B_0,V,T)}^2}{\overline{
G^2(B_0,V,T)}-\overline{G(B_0,V,T)}^2},
\]
where over-line indicates averaging over magnetic field interval
$0<B_0<10T$.  Fig.~\ref{correl}-A displays the field
autocorrelation function at $V=0$ at base temperature.

Next, we average the correlation field among different thermal
cycles. The resulting correlation field versus bias voltage at
base temperature is displayed in Fig.~\ref{correl}-B. Clearly,
$B_C(V,T)$ increases with $V$. In voltage interval $0<V<200\mu V$,
the voltage dependence is approximately linear, and the best
linear fit is
\[
B_C(V,T)=0.28+4724V,
\]
where the left hand side is in Tesla and $V$ is in volt.

As discussed above, the voltage dependence of the correlation
field displays the energy dependence of the dephasing time.
Substituting $E=eV/2$ for the average electron energy injected
from the Fermi level of the source, we find

\begin{equation}
\mu_B B_C(V,T)=32\mu eV+0.98\cdot E. \label{bcdata}
\end{equation}

At energy $E\gg 32\mu eV$, the offset in this equation can be
neglected. Similarly, at large energy ($E\gg k_B T$), the offset
in Eq.~\ref{bc} can be neglected, so comparing Eq.~\ref{bc} and
Eq.~\ref{bcdata}, we find
\begin{equation}
\frac{h}{\tau_\phi (E,T)}=0.98E. \label{deph}
\end{equation}

It follows that the correlation energy of an electron with energy
$E$ above the Fermi level is equal to $E$. Before we discuss the
physical meaning of this observation, we examine the correlation
scale at zero bias voltage.

At $E=eV/2=0$, the correlation energy is $E_C (0,T)=\mu_B
B_C(0,T)= 32\mu eV$. This value is reasonable if we assume that
the base electron temperature is $T=57mK$. In particular, at
$V=0$, the average electronic energy is given by the
full-width-half-maximum of the derivative of the Fermi function
($3.5k_BT$). So, $E_C(0,T)\sim 2.9k_BT+0.98\cdot 3.5k_BT=6.4k_BT$.
Substituting $T=57mK$, we obtain $E_C=32\mu eV$. $T=57mK$ is a
reasonable value for our base electron temperature, because the
wires were filtered at the base phonon temperature ($15mK$).

We believe that the equation $h/\tau_\phi (E)= E$ demonstrates a
break-down of the Fermi-Liquid description of electron transport
in our samples. The selected samples have conductance comparable
to the conductance quantum. Coulomb Blockade in these samples is
barely suppressed and the localization length is comparable to
sample size. We expect that the Fermi-liquid picture in this
regime begins to fail, which causes the energy uncertainty of a
quasiparticle with energy $E$ to be comparable to $E$. So, the
correlation energy is comparable to the excitation energy.

Now we discuss the variation of the Thouless energy among devices
with a similar resistance. The Thouless energy is directly related
to the electron diffusion time across the device, which implies
that the electron diffusion time varies among different samples.
The diffusion time is shorter than the thermal time ($h/k_B T$) in
two out of ten samples with resistance slightly below the
resistance quantum.

The conductance-fluctuation amplitude is comparable to $e^2/h$ in
all ten samples. This implies that in the eight samples where the
Thouless energy is $<k_BT$, the Thouless energy cannot be $\ll
k_BT$. If we assume the opposite, then the conductance-fluctuation
amplitude will be suppressed by $\sqrt{E_{Th}/k_B T}\ll
1$,~\cite{lee1} contrary to the data. So, the diffusion time in
our samples varies roughly within an order of magnitude.

The fluctuations in the diffusion time among samples with similar
resistance can be explained by the fluctuations in sample
dimensions $w$ and $L$. The fabrication process is based on
electromigration and we do not have precise control over these
dimensions. We found through scanning electron microscopy that the
conductance is only roughly proportional to the apparent width w.
The conductance per unit width varies within a factor of 3 among
different samples. Since the diffusion time scales as $L^2$ and
the selected ten samples have similar resistance, the diffusion
time can vary by factor of $\sim 9$, even if we assume that the
resistivity does not vary among samples. This can explain the
observed variation in the Thouless energy among samples.

The fact that the fluctuation in the diffusion time among samples
is within an order of magnitude indicates that our sample
parameter control is also within an order of magnitude. We think
that this sample control is quite good, in the light that the
nanojunctions were self-created through electromigration.

\section{Appendix: Derivation of the Correlation Field at finite temperature}

At finite temperature, Eq.~\ref{conductance} still applies, but
with $f$ replaced with the Fermi function at temperature $T$. The
formula takes into account thermal broadening effects, but it
neglects dephasing.

First we account for the thermal broadening. In
Eq.~\ref{conductance0}, which is valid at $T=0$, we found that the
correlation voltage of the integral is three times larger than
that for the first term on the right hand side of the equation.
Thus, to find the temperature dependence $V_C(T)$ in the limit of
small $T$, we only need to find the correlation voltage of the
first integral in Eq.~\ref{conductance}. The effect of the second
integral in Eq.~\ref{conductance} on $V_C(T)$ is weak if
$V_C(T)<3V_C(0)=6E_{Th}$.

If we neglect the integrals in Eqs.~\ref{conductance0} and
~\ref{conductance}, the conductance becomes
\[
\tilde G(B,V,0)=\frac{e^2}{2h}\sum_{\alpha,\alpha'=\pm 1}
T(\alpha\mu_B B+ \alpha'\frac{eV}{2},V)
\]
at $T=0$ and
\begin{widetext}
\[
\tilde G(B,V,T)=-\sum_{\alpha,\alpha'=\pm 1}\frac{e^2}{2h}\int
dE\cdot T(E+\alpha\mu_BB,V)f'(E+\alpha'\frac{eV}{2})
\]
\end{widetext}
at $T>0$. It follows that
\begin{equation}
\tilde G (B,V,T)=\int dx (-f'(x)) \tilde G (x/\mu_B+B,V,0).
\label{convolve}
\end{equation}

In this approximation, the conductance fluctuations at $T>0$ can
be obtained by convolving the conductance fluctuations at $T=0$
with the derivative of the Fermi function. We reiterate that we
neglected the conductance contributions with larger correlation
scales.

It follows that the field correlation function at temperature $T$
is proportional to
\[
\int\int dxdy f'(x)f'(y)\tilde C((x-y)/\mu_B+B,0) .
\]
where $\tilde C$ is the field correlation function at $T=0$.  We
can make an approximation $\tilde
C((x-y)/\mu_B+B,0)\sim\delta((x-y)/\mu_B+B)$, because the thermal
of $f'$ ($3.5k_B T$) is much larger the width of $\tilde
C((x-y)/\mu_B+B,0)$ (the Thouless energy).  So, the field
correlation function at temperature $T$ becomes proportional to
the autocorrelation function of the derivative of the Fermi
function. The correlation field $\mu_B B_C$ is the
half-width-half-maximum of the autocorrelation function of $f'$
(which is $2.9k_B T$), so
\[
\mu_B B_C=E_C=2.9k_BT.
\]

In the presence of dephasing, the correlation energy is increased
by energy uncertainty $h/\tau_\phi$.~\cite{datta}. So, the
correlation field becomes
\[
\mu_B B_C=E_C=2.9k_BT+\frac{h}{\tau_\phi}.
\]
In good Fermi liquids, one can neglect the dephasing term.
\section{conclusion}

We show experimentally and theoretically that quantum interference
effects with magnetic field in highly disordered Au nanobridges
are governed by the Zeeman effect. The Aharonov-Bohm effect is
suppressed when the nominal Drude mean free path is smaller than
the Fermi wavelength. Using the Landauer-Buttiker formalism, we
show that both magnetofingerprints and  conductance fluctuations
with bias voltage are equivalent to the Ericson's fluctuations.
This permits measurements of the correlation energy from the
correlation field. If sample size is comparable to the
localization length, then the correlation energy of electrons at
energy $E\gg k_BT$ will be $\approx E$, which shows that the Fermi
liquid description breaks down at the localization length-scale.

This work was performed in part at the Cornell Nanofabrication
Facility, (a member of the National Nanofabrication Users
Network), which is supported by the NSF, under grant ECS-9731293,
Cornell University and Industrial affiliates, and the Georgia-Tech
electron microscopy facility. This research is supported by the
David and Lucile Packard Foundation grant 2000-13874 and the NSF
grant DMR-0102960.

\bibliography{career1}
\end{document}